\documentclass[12pt,preprint]{aastex}

\begin{document}

\newcommand\etal{et al. }

\def\t0{\theta_{\circ}}
\def\muo{\mu_{\circ}}
\def\sd{\partial}
\def\be{\begin{equation}}
\def\en{\end{equation}}
\def\bv{\bf v}
\def\bvo{\bf v_{\circ}}
\def\ro{r_{\circ}}
\def\rhoo{\rho_{\circ}}
\def\etal{et al.\ }
\def\msun{M_{\sun}}
\def\rsun{R_{\sun}}
\def\lsun{L_{\sun}}
\def\msunyr{M_{\sun} \, yr^{-1}}
\def\kms{\rm \, km \, s^{-1}}
\def\mdot{\dot{M}}
\def\ha{H$\alpha \;$}
\def\ecs{\rm erg \, cm^{-2} \, s^{-1}}

\title{Unveiling the Inner Disk Structure of T Tauri Stars}
\author{James Muzerolle \altaffilmark{1}, Nuria Calvet \altaffilmark{2},
Lee Hartmann \altaffilmark{2}, and Paola D'Alessio\altaffilmark{3}}
\altaffiltext{1}{Steward Observatory, University of Arizona, 933 N. Cherry
Ave., Tucson, AZ 85721 (jamesm@as.arizona.edu)}
\altaffiltext{2}{Harvard-Smithsonian Center for Astrophysics, Mail Stop 42,
60 Garden Street, Cambridge, MA 02138 (ncalvet@cfa.harvard.edu, hartmann@cfa.harvard.edu)}
\altaffiltext{3}{Instituto de Astronom\'{\i}a, UNAM, Ap. Postal
3-72, 58089 Morelia, M\'exico (p.dalessio@astrosmo.unam.mx)}

\begin{abstract}

We present near-infrared spectra of the excess continuum emission from
the innermost regions of classical T Tauri disks.  In almost all cases,
the shape of the excess is consistent with that of a single-temperature
blackbody with $T \sim 1400$ K, similar to the expected dust sublimation
temperature for typical dust compositions.  The amount of excess flux 
roughly correlates with the accretion luminosity in objects with similar stellar
properties.  We compare our observations with the predictions of
simple disk models having an inner rim located at the dust
sublimation radius, including irradiation heating of the dust from both
the stellar {\it and} accretion luminosities.  The models
yield inner rim radii in the range 0.07-0.54 AU, increasing
with higher stellar and accretion luminosities.  Using typical parameters
which fit our observed sample, we predict a rim radius $\sim 0.2$ AU
for the T Tauri star DG Tau, which agrees
with recent Keck near-infrared interferometric measurements.
For large mass accretion rates, the inner rim lies beyond
the corotation radius at (or within) which magnetospheric accretion
flows are launched, which implies that pure
gaseous disks must extend inside the dust rim.  
Thus, for a significant fraction of young stars, dust cannot exist
in the innermost disk,
calling into question theories in which solid particles are ejected
by a wind originating at the magnetospheric radius.
\end{abstract}


\section{Introduction}

The structure of inner disk regions of classical T Tauri stars (CTTSs)
has important consequences for both the transfer of material to the star
and the processing of solid material in the terrestrial planet region.
It is now generally accepted that the accretion of disk material onto the
central star is channeled by the stellar magnetosphere. The inner edge of the disk
is truncated by the stellar magnetic field at or inside the corotation radius,
where the disk rotates at the same angular velocity as the star
(e.g., Shu et al.  1994), typically a few stellar radii from the stellar
surface.  Gas from the disk
is channeled out of the disk plane along magnetic field lines to impact
the stellar surface in an accretion shock.  Observations and modeling of both
permitted emission line profiles produced in the infalling gas streams
(Muzerolle, Hartmann, \& Calvet 2001, and references therein) and hot
continuum excess produced by the accretion shock (Calvet \& Gullbring 1998)
lend strong support to this picture.

However, substantial uncertainty remains concerning the detailed structure of
the disk near the inner truncation radius.  Meyer, Calvet, \& Hillenbrand
(1997: MCH97) showed that the near-infrared excess emission of CTTSs,
which arises from inner disk regions, is correlated with accretion rate.
However, quantitative agreement with standard disk models
required accretion rates higher by a factor of $\sim$ 10 than subsequently 
estimated from UV excess emission (e.g., Gullbring et al. 1998).
Recent determinations of near-infrared continuum excesses or ``veiling'' 
also showed larger disk emission than predicted by simple models
(Folha \& Emerson 1999); Johns-Krull \& Valenti (2001) found that
disk models such as those of Chiang \& Goldreich (1997, 1999)
and D'Alessio et al. (1998)
underestimated the $K$-band veiling by factors of up to $\sim$ 2-3.

The higher mass Herbig Ae/Be stars (HAEBESs) also show large near-infrared
excesses difficult to explain with simple accretion disk models
(Hillenbrand et al. 1992; Hartmann, Kenyon, \& Calvet 1993).
Natta et al. (2001) attempted to solve this problem by postulating 
that the near-infrared emission arises from an inner disk rim at the dust
destruction radius.
The rim is ``puffed'' because of the normal incidence of the
stellar radiation, which is the primary source of dust heating.
At the same time, Tuthill et al. (2001) independently proposed
dust sublimation to describe the near-infrared interferometric size
for the HAEBES Lk\ha 101.
Other recent near-infrared interferometric observations of
HAEBESs have resolved structures with
sizes and visibilities in qualitative agreement with the puffed rim picture
(Millan-Gabet, Schloerb, \& Traub 2001; Eisner et al. 2003).
A similar inner dust rim in CTTSs seems likely; an inner edge must already
exist near the corotation radius by magnetic truncation, and dust cannot
survive inside this radius.

To investigate this possibility, we have obtained 2-5 $\mu$m spectra of
a sample of well-studied CTTSs.  These spectra, with simultaneous coverage
over a wide wavelength range, provide a far more sensitive
probe of the shape and strength of the excess emission
than previous photometric measurements.
Unlike HAEBESs, the near-infared continuum excesses of CTTSs
do not stand out strongly in contrast with the stellar photosphere.
The only way to disentangle the two broad continua (with less dissimilar
characteristic temperatures) is to determine the spectrum of the excess
by measuring absorption line veiling from high-quality spectra simultaneously
spanning a large range of wavelength.
We show that the excess spectra derived from our data appear to be
dominated by a single-temperature blackbody
component with characteristic temperature roughly corresponding to
the dust sublimation limit.  We then compute models for an inner dust rim,
from which we can estimate the dust truncation radii, $R_d$.

\section{Spectra of the Infrared Excess}

We observed a sample of 9 CTTSs with SpeX on IRTF
(Rayner et al. 1998) during the nights of January 5-6, 2001.
Each spectrum we observed spans a wavelength range of about 2.1-4.8 $\mu$m,
which contains multiple hydrogen emission lines that serve as key
accretion diagnostics (see \S 3), and $K$-band atomic
absorption features that can be used to measure continuum veiling
of the stellar photosphere from disk emission.
We selected a representative sample of pre-main sequence objects,
spanning a wide range of stellar and accretion properties.
These objects have been studied extensively, and the stellar
parameters are well-determined (see Table 1), with the exception of
DR Tau, the most active object with the largest amount of optical
continuum excess.

Most of the basic data reduction steps were performed using
the facility IDL extraction software ``Spextool" (Cushing et al. 2003).
Removal of telluric absorption lines
was done separately; each object spectrum was divided by
a solar-type dwarf standard spectrum taken at a similar airmass
and sky position, and then multiplied by the solar spectrum to remove
residuals from photospheric features in the standard spectrum.
Spectra from each order (6 orders in all) were then pieced together
into one final spectrum for each object.
The overall continuum level across all the orders is remarkably
consistent, with generally good agreement in the regions of order overlap.

We measured the veiling at 2.2 $\mu$m ($r_{2.2} = F_{excess,2.2}/F_{*,2.2}$)
by comparing each object spectrum with a weak T Tauri or dwarf template spectrum
of similar spectral type, after normalizing all the spectra by the observed
flux at the shortest wavelength (roughly 2.12 $\mu$m).  The template spectra
were then artificially veiled by adding an excess continuum until the depth
of photospheric features around 2.2 $\mu$m matched the depth of the features
in the object spectra.  The measured veilings are listed in Table 1.
These are in relative agreement with previous determinations
from high-resolution spectra
(Folha \& Emerson 1999; Johns-Krull \& Valenti 2001).
Since the data are not absolutely flux calibrated, we used the veiling values
to scale the object spectra relative to the unveiled template spectra.
Both object and template spectra were dereddened using published $A_V$ values.
Finally, excess spectra were created by subtracting each dereddened,
veiling-scaled object spectrum by the appropriate dereddened template spectrum.

\begin{figure}
\plotone{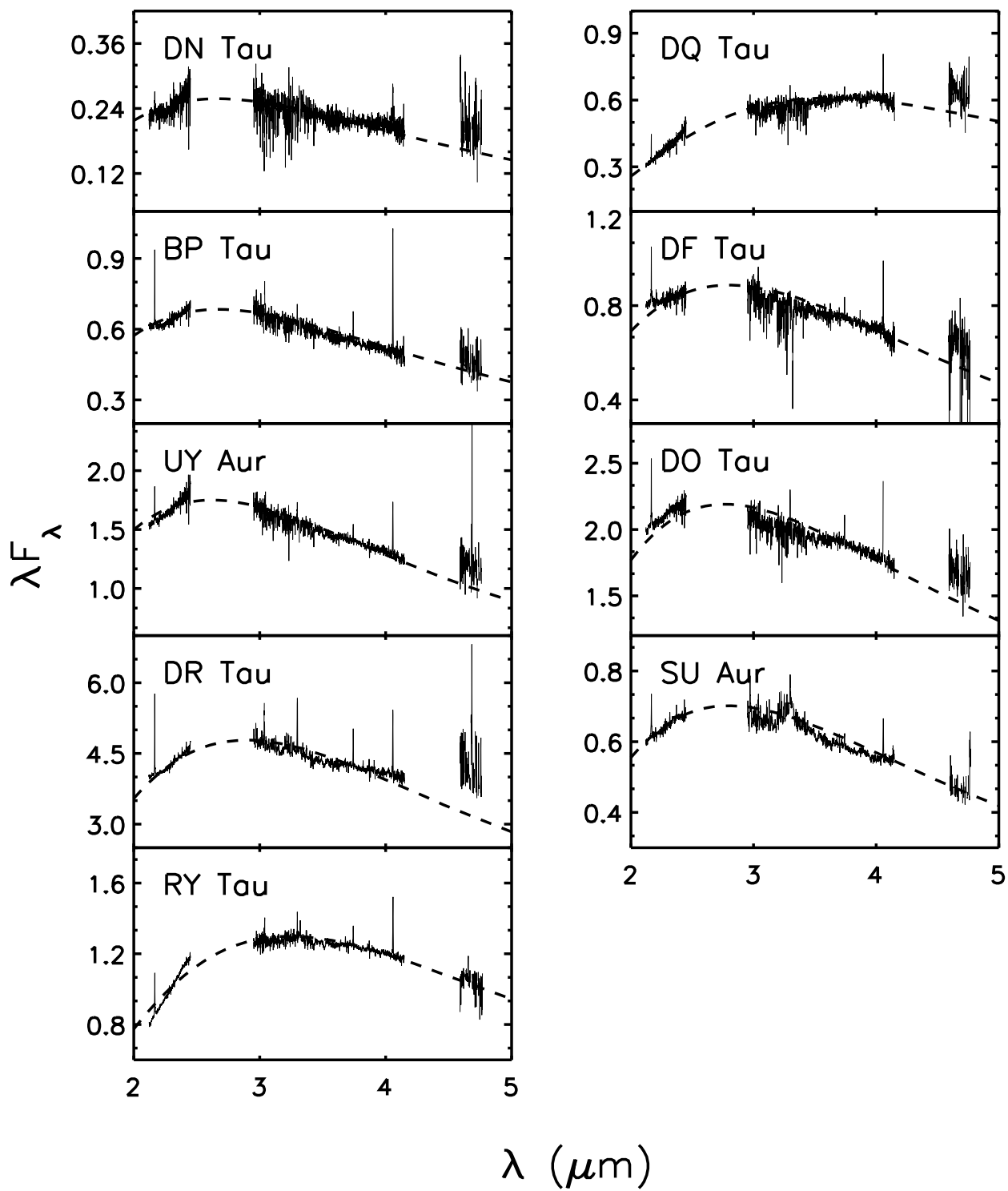}
\caption{Near-infrared excess spectra of classical T Tauri stars.
Each dereddened object spectrum has been subtracted by that of
a dereddened weak T Tauri star
of the same spectral type (with no excess).  Apparent absorption features
in the $L$ and $M$ bands are residuals from incomplete telluric cancellation.
Prominent Brackett and Pfund emission lines, produced by accretion flows,
can be seen in most of the spectra.
The dashed lines show the emission from the best-fit inner dust rim models.
Fluxes are in units of the photospheric flux at 2.1 $\mu$m.
\label{excess}}
\end{figure}

The resultant excess spectra are shown in Figure~\ref{excess}.
The excess flux appears to increase with wavelength through the $K$-band,
but turns over between 2.5 - 3 $\mu$m (with the exception of DQ Tau,
where the flux increases with wavelength throughout the observed range).
The spectral shape implies emission from optically thick material at
a characteristic temperature; we were able to qualitatively match the shapes
using single-temperature blackbodies with $T \sim 1200 - 1400 \; K$.
These temperatures are close to the dust sublimation limit.
We next explore models of emission from an accretion disk with a puffed
inner rim corresponding to the dust destruction radius.

\section{Models}

Our model of the inner disk rim will be presented in more detail
in a forthcoming paper; see also D'Alessio et al. (2003) for further
explanation of the radiative transfer, as well as other comparisons
with observed spectral energy distributions.  Here, we briefly summarize
the main points of the model.
We assume that a star of effective temperature, radius, mass,
and luminosity given by $T_*$, $R_*$, $M_*$, and $L_*$
is surrounded by a disk, accreting at a rate $\mdot$,
which has a rim at the radius where dust sublimates.
For simplicity, we assume this rim to be vertical,
even though radial and vertical gradients of density
and temperature probably result in a more rounded shape.
Our models include a rim atmosphere allowing for the transition
from optically-thin to optically-thick absorption and emission.
The rim is illuminated by both the star and
by radiation emerging from the accretion shock
where most of the accretion luminosity $L_{acc} = G M_* \mdot /R_*$
is dissipated (Calvet \& Gullbring 1998).

We follow the treatment of Calvet et al. (1991, 1992=CMPD)
to describe the temperature structure in the
rim atmosphere, extended to include scattering of radiation
at the disk characteristic wavelength;  we also use the forward
scattering approximation, which amounts to taking negligible
albedo in the CMPD treatment.
The temperature at the top of the
rim atmosphere, located at radius $R$, is given by
\begin{equation}
T(0)^4 = { E_0 \over { 4 \sigma_R}} \left ( 2 + q  {\chi_d \over \kappa_d} \right  )
\end{equation}
where $E_0 = (L_* + L_{acc})/4 \pi R^2$,
and the temperature at the level $\tau_d \sim 2/3$ is
\begin{equation}
T(2/3)^4 = T_{eff}^4 = { E_0 \over { 4 \sigma_R}}
\left [ 2 + {3 \over q} + (q {\chi_d \over \kappa_d} - {3 \over q}) e^{-2 q / 3 } \right ].
\label{teff}
\end{equation}
In these equations, $q$ is an efficiency factor, given
by the ratio of opacities at wavelengths characteristic
of the absorption and emission of radiation by the
dust in the rim (CMPD), i.e., $q = \bar \kappa^i / \chi^d$,
where $\bar \kappa^i = (L_* \kappa^* + L_{acc} \kappa^{acc})/(L_* + L_{acc})$.
In these equations, $\chi$ refers to the total opacity and 
$\kappa$ to the true opacity, each at a wavelength characteristic
of the dust in the disk rim (subindex $d$) and of the incident radiation
(subindex $i$), in turn composed of stellar ($\kappa^*$)
and accretion shock ($\kappa^{acc}$) radiation.  In practice,
 $\chi$ and $\kappa$ are means of the monochromatic opacity weighted
by blackbodies at $T_d$, $T_*$, and $T_{acc}$, for the three
characteristic radiation fields. We take $T_{acc} = $ 8000 K
(Calvet \& Gullbring 1998).  The rim is then located at radius
\begin{equation}
R_d = \left [ \left ( { {L_* + L_{acc}} \over { 16 \pi \sigma_R}} \right ) \,\,  (2 +q {\chi_d \over \kappa_d}) \right ]^{1/2}
 { 1 \over T_d^2}
\label{rd}
\end{equation}
where we have assumed that the thickness of the atmosphere
is negligible compared to the radius.

We calculate the scale height of the rim, $H$,
using eq.(\ref{teff}).  The physical height
of the rim, $z_{rim}$, is then calculated using
the expression in Muzerolle et al. (2003) with
a central density given by $\rho_c \sim \Sigma / \sqrt{2 \pi}  H$,
where the surface density $\Sigma = \mdot / 6 \pi\nu [( 1 - (R_*/R_d)^{1/2}]$, 
and the mass accretion rate $\mdot$ is derived from $L_{acc}$.
The viscosity is given by $\nu = \alpha c_s(T_{eff})^2/ \Omega_K(R_d)$,
where $c_s$ is the sound velocity and $\Omega_K$ is the Keplerian
angular velocity, and we adopt $\alpha = 0.01$,
as suggested by our SED modeling of CTTS disks (D'Alessio et al. 2001).
Finally, the inclination-dependent flux of the
rim is calculated as $F_{\lambda} = I_{\lambda} A(i)$,
where the rim area $A(i)$ has been derived from expressions
in Appendix B in Dullemond et al. (2001). The specific
intensity $I_{\lambda}$ is calculated by integrating the
temperature profile of the atmosphere (CMPD); in practice,
$I_{\lambda}$ differs from $B_{\lambda}(T_{eff})$ by only a few percent.  
The location and height of the rim are highly dependent on the dust
properties.  To describe the dust opacity we take silicate abundances
from Pollack et al. (1994), and a dust size distribution
given by $n(a) \propto a^{-3.5}$ between $a_{min} = 0.005 \mu$m
and a variable $a_{max}$ (see D'Alessio et al. 2001 for details).

Given the stellar properties and $L_{acc}$, the
emission from the rim depends on three parameters:
the dust opacity, the dust destruction temperature, and
the inclination.  We have measured 
the {\em simultaneous} $L_{acc}$ (i.e., derived from
the same data as was the excess spectrum) for each object
from the flux of the Br$\gamma$ line in each spectrum,
employing the calibration of Muzerolle, Hartmann, \& Calvet (1998).
These values of $L_{acc}$ are given in Table 1.
The model parameters for $R_d$, $T_d$, $a_{max}$, $i$, and $q$
that provide the best fit to the dereddened excesses are also given
in Table 1; the emission from these models 
is shown in Figure 1. In general, the agreement between
model and observation is excellent. The largest departure
is observed at the longest wavelength for the 
object with one of the largest mass accretion rates in the sample
(i.e. DR Tau).  This is likely due to a contribution to the flux from
the inner disk just outside the rim, which has not been included
in our modeling.  A few other objects show slightly larger fluxes
than the models at long wavelengths, which in most cases is also
likely due to emission from the disk just outside the rim.
However, we note that the telluric absorption is especially strong
in the $M$ band, thus the flux discrepancies in some cases
may also be due to uncertainties in the telluric cancellation.

The fits require small grains, $\le 1 \mu$m, in the atmosphere of the
rim.  This result is surprising in that much larger
values of $a_{max}$ are required to fit the
median SED in Taurus (D'Alessio et al 2001), in agreement
with flat slopes in the millimeter range. However,
this may be an indication that dust settling has 
already taken place in the inner disk. Weidenschilling (1997)
found that big grains settle to the disk midplane first, leaving 
behind a population of small grains.

The dust temperatures ($T_d$) depend only on the shape of the excess
spectra; most of the objects have $T_d \sim 1400$ K.
Sublimation temperatures
of $T_d \lesssim 1400$ K are expected given the central densities
at the rim $\lesssim 10^{-8}$ g cm$^{-3}$ (Pollack et al. 1994).
We also find that $T(2/3)$ is within $\sim 100$ K from $T_d$,
indicating that the rim atmosphere is nearly isothermal.
Moreover, the rim heights are typically $\sim 0.01 - 0.1$ AU
and the ratio of the rim height to the scale height in
all cases is $\sim 4$, in agreement with values for HAEBESs.
(Dullemond et al. 2001).
Note that our models are qualitatively similar to those of
Natta et al. (2001) and Dullemond et al. (2001), but differ in
several key respects.  One, we include an atmosphere for the rim.
Two, we take into account the wavelength dependence of the incident
radiation field and the dust opacity (as in CMPD and Monnier \&
Millan-Gabet 2002).  Three, and most importantly,
we include $L_{acc}$ as a heating source; inclusion of $L_{acc}$
has important consquences for predicting the rim radius, as we discuss
in the next section.

\section{Discussion}

Our models of irradiated inner disk rims account for the large
near-infrared excesses of CTTSs without requiring accretion rates that
are too large.  At the same time, the models explain the correlation 
between near-infrared excess and accretion rate found by MCH97, because the
accretion shock radiation effectively increases the rim radiation.

Our models make several predictions that can be tested by further 
observations.  One of these is the inner radius of the dust disk;  
the best-fit values for each object are shown in Table 2.
There is a two-fold increase of the rim radius (0.07 to 0.15 AU)
for a factor of $\sim 100$ increase in $L_{acc}$\footnote{The dependence on
accretion does not hold for DQ Tau, which
has a much larger $R_d$ than expected from its $L_{acc}$.
This object is a spectroscopic binary with semi-major axis
$a \sim$ 0.067 AU (Mathieu et al. 1997).  If the
circumbinary disk is truncated at $\sim 2 a$ (Artymowicz \&
Lubow 1994), then the truncation radius would be 0.13 AU,
which is equal to our estimate for $R_d$.
This suggests that the near infrared excess in DQ Tau
corresponds to emission from the inner edge of the
circumbinary disk.}. 
The two most luminous CTTSs in the sample, SU Aur and RY Tau,
also have significantly larger rim radii (0.36 and 0.54 AU, respectively).
In many cases the predicted inner rim radii produce near-infrared emission
on spatial scales within reach of the capabilities of current interferometers.
In fact, Akeson et al. (2002) estimated a size scale for the $K$-band emission
from SU Aur of a few tenths of an AU, similar to our predicted value.
Furthermore, just-published Keck interferometric observations of
the CTTS DG Tau (Colavita et al. 2003) resolved the $K$-band emission
and derived a radius of 0.12-0.24 AU.  Using model parameters typical of
the fits for our sample ($q=1$, ${\chi}_d/{\kappa}_d=2$, $T_d=1400 \; K$),
and the known parameters of DG Tau ($L_*=1.07 \; L_{\odot}$,
$L_{acc}=1.12 \; L_{\odot}$), we predict a rim radius of $\sim$0.2 AU,
in excellent agreement with the Keck measurements.
As we have already mentioned, models of dust sublimation
have been been used
before in trying to explain interferometric observations of HAEBESs
(Tuthill et al. 2001; Natta et al. 2001; Dullemond et al. 2001;
Monnier \& Millan-Gabet 2002).  However, these models
have not included the contribution from $L_{acc}$, and thus are likely
to underestimate the emission sizes of most CTTSs.
Future interferometric observations of CTTSs of
a wide range of accretion and stellar luminosities should be
able to confirm the relationships implied by our results.

Stars with the lowest $L_{acc}$ have $R_d$ close to or slightly larger 
than their corotation radii (see Table 1); however, as $L_{acc}$ increases,
$R_d$ becomes much larger.  Since all of these stars are accreting gas,
and accretion can only occur if disk material extends all the way to or
inside of the corotation radius, this then implies that stars with high
$L_{acc}$ have a dust-free pure gaseous innermost disk.
Such a scenario is similar to our recent investigation
of Herbig Ae stars (Muzerolle et al. 2003), where we combined
a puffed inner dust rim with the D'Alessio et al. (1997)
detailed accretion disk structure, using accretion rates estimated from
UV fluxes or permitted line profile modeling.
Because of the typically low values
of the HAEBES mass accretion rates ($\sim 10^{-8} \msunyr$),
the gas disk is either optically thin or else the height where the optical
depth to the stellar radiation is unity is significantly reduced
due to the lack of dust.  This is also true for the CTTSs.
Thus, the inner dust rim can still receive
the direct stellar irradiation necessary to maintain its enhanced
scale height.  In the case of the hotter CTTSs presented here,
the dependence of the inner dust rim radius on accretion
is negligible since $L_* > L_{acc}$; however, there must still be
an inner gaseous disk to feed the accretion flows onto these stars.

The inner dust rim model successfully accounts for all of the continuum
excess veiling observed at $K$ in CTTSs.  This model may also explain
the large near-infrared veiling observed in many protostars
(Casali \& Eiroa 1996; Greene \& Lada 1996), as these objects
probably have higher accretion
rates (Muzerolle \etal 1998; Greene \& Lada 2002).

Finally, the presence of a dust-free gas disk inside $R_d$ in many CTTSs 
suggests that there may be difficulties for the X-wind model of
chondrule formation (e.g., Shu \etal 2001), which invokes a wind arising from
just outside the magnetosphere to lift solid particles out of the disk
and carry them out to larger radii. If dust cannot exist in this region, 
which our results indicate is the case for the rapid accretion phases
required by the X-wind model, then an alternative theory may
be required to explain chondrule formation.

\acknowledgements

JM would like to thank John Rayner and Bobby Bus at IRTF for their able 
assistance with the SpeX observations and data reduction.
This work was supported in part by NASA Origins of Solar Systems
grant NAG5-9670.


\clearpage

\begin{deluxetable}{lccccccccccccc}
\tabletypesize{\scriptsize}
\tablecaption{CTTS Sample \& Model Parameters\label{tab1}}
\tablehead{
\colhead{Object} & \colhead{Sp. Type} & \colhead{$L_*$} &
\colhead{$R_*$} & \colhead{$A_V$} & \colhead{$i_*$} & \colhead{$R_{co}$} &
\colhead{$r_{2.2}$} & \colhead{$L_{acc}$} & \colhead{$R_d$} &
\colhead{$a_{max}$} & \colhead{$T_d$} & \colhead{$i$} & \colhead{$q$}\\
\colhead{} & \colhead{} & \colhead{($L_{\odot}$)} & \colhead{($R_{\odot}$)} &
\colhead{} & \colhead{($^\circ$)} & \colhead{(AU)} & \colhead{} &
\colhead{($L_{\odot}$)} & \colhead{(AU)} & \colhead{($\mu$m)} & \colhead{(K)} &
\colhead{($^\circ$)} & \colhead{}}
\startdata
DN Tau & M0 & 0.86 & 2.1 & 0.25 & 35 & 0.05 & 0.2 $\pm$ 0.2 & $<$0.01 & 0.07 & 0.1 & 1400 & 28 & 3.24\\
DQ Tau & M0 & 0.63 & 1.8 & 0.97 & 23$^a$ & \nodata & 0.3 $\pm$ 0.2 & 0.04 & 0.13 & 0.3 & 1000 & 17 & 1.39\\
BP Tau & K7 & 0.87 & 1.9 & 0.51 & 39 & 0.07 & 0.6 $\pm$ 0.2 & 0.15 & 0.09 & 0.2
& 1400 & 26 & 1.65\\
DF Tau & M1 & 0.5 & 1.7 & 0.45 & \nodata & 0.07 & 0.8 $\pm$ 0.3 & 0.22 & 0.09 &
0.7 & 1300 & 61 & 0.98\\
UY Aur & K7 & 1.07 & 2.1 & 1.26 & 42$^b$ & \nodata & 1.5 $\pm$ 0.5 & 0.19 & 0.1
& 0.3 & 1400 & 67 & 1.31\\
DO Tau & M0 & 1.03 & 2.3 & 2.27 & \nodata & \nodata & 2.0 $\pm$ 0.5 & 0.18 & 0.09 & 0.1 & 1400 & 63 & 2.55\\
DR Tau & K7: & 0.87: & 1.9: & 1.6 & 69 & 0.08 & 4.0 $\pm$ 0.5 & 1.01 & 0.15 & 0.5 & 1300 & 72 & 1.56\\
SU Aur & G2 & 12.9 & 3.5 & 0.9 & 72 & \nodata & 0.6 $\pm$ 0.3 & 1.47 & 0.36 & 0.2 & 1400 & 86 & 2.21\\
RY Tau & G5 & 12.8 & 3.6 & 1.8 & \nodata & \nodata & 0.8 $\pm$ 0.3 & 4.04 & 0.54 & 0.3 & 1200 & 86 & 2.06
\tablecomments{The stellar inclination angles, $i_*$, were estimated from
$v\sin i$ (Hartmann \& Stauffer 1989), $R_*$, and the period
(Johns-Krull \& Gafford 2002, and references therein; Bouvier et al. 1993),
except for: (a) taken from the Mathieu et al. (1997) estimate based on
orbital components of the spectroscopic binary; (b) taken from
the Close et al. (1998) estimate of the inclination of the circumbinary disk.
Colons after the DR Tau stellar parameters indicate large uncertainty,
due to the extremely strong optical veiling.
$R_{co}$ is the corotation radius, estimated from the stellar rotation rates.
The rim model inclination angles, $i$, have uncertainties
of $\sim \pm 10^{\circ}$.}
\enddata
\end{deluxetable}

\clearpage

\end{document}